\documentclass[letterpaper]{article} 
\usepackage{aaai25}  
\usepackage{times}  
\usepackage{helvet}  
\usepackage{courier}  
\usepackage[hyphens]{url}  
\usepackage{graphicx} 
\usepackage{amsmath}
\usepackage{textcomp}
\usepackage{booktabs}
\usepackage{natbib}  
\usepackage{caption} 
\usepackage{algorithm}
\usepackage{algorithmic}
\usepackage{newfloat}
\usepackage{listings}
\DeclareCaptionStyle{ruled}{labelfont=normalfont,labelsep=colon,strut=off} 
\floatstyle{ruled}
\newfloat{listing}{tb}{lst}{}
\floatname{listing}{Listing}
\usepackage{amsmath}   
\usepackage{amssymb}   

\title{Unpaired Multi-Domain Histopathology Virtual Staining using Dual Path Prompted Inversion } 
\author{
    Bing Xiong\textsuperscript{\rm 1,2},
    Yue Peng\textsuperscript{\rm 1,2},
    RanRan Zhang\textsuperscript{\rm 1,2},
    Fuqiang Chen\textsuperscript{\rm 1,2},
    JiaYe He\textsuperscript{\rm 1,3},
    Wenjian Qin\thanks{*Corresponding author}\textsuperscript{\rm 1,2}
}
\affiliations{
    \textsuperscript{\rm 1}Shenzhen Institutes of Advanced Technology, Chinese Academy of Sciences, SheZhen 518000,China\\

    \textsuperscript{\rm 2}University of Chinese Academy of Sciences, Beijing 100000,China\\

    \textsuperscript{\rm 3}National innovation center for advanced medical devices, SheZhen 518000,China\\
}

\usepackage{bibentry}

\begin{document}

\maketitle

\begin{abstract}
Virtual staining leverages computer-aided techniques to transfer the style of histochemically stained tissue samples to other staining types. In virtual staining of pathological images, maintaining strict structural consistency is crucial, as these images emphasize structural integrity more than natural images. Even slight structural alterations can lead to deviations in diagnostic semantic information. Furthermore, the unpaired characteristic of virtual staining data may compromise the preservation of pathological diagnostic content. To address these challenges, we propose a dual-path inversion virtual staining method using prompt learning, which optimizes visual prompts to control content and style, while preserving complete pathological diagnostic content. Our proposed inversion technique comprises two key components: (1) \textit{\textbf{Dual Path Prompted Strategy}}, we utilize a feature adapter function to generate reference images for inversion, providing style templates for input image inversion, called  Style Target Path. We utilize the inversion of the input image as the Structural Target path, employing visual prompt images to maintain structural consistency in this path while preserving style information from the style Target path. During the deterministic sampling process, we achieve complete content-style disentanglement through a plug-and-play embedding visual prompt approach.  (2) \textit{\textbf{StainPrompt Optimization}}, where we only optimize the null visual prompt as ``operator'' for dual path inversion, rather than fine-tune pre-trained model. We optimize null visual prompt for structual and style trajectory around pivotal noise on each timestep,  ensuring accurate dual-path inversion reconstruction. Extensive evaluations on publicly available multi-domain unpaired staining datasets demonstrate high structural consistency and accurate style transfer results.Our code and Supplementary materials are available
at:\url{https://github.com/DianaNerualNetwork/StainPromptInversion}

\end{abstract}

\section{Introduction}

Histopathological examination is widely regarded as the clinical gold standard for disease diagnosis. This process often involves histochemical staining, where different tissue components are differentiated by distinct colors to aid pathological diagnosis. However, traditional histochemical staining can cause color interference when re-staining a sample, making observation difficult. This constraint hampers their broader application in histopathology. Routine pathological examinations typically employ hematoxylin and eosin (H\&E) staining to highlight tissue morphology for preliminary diagnosis. Nevertheless, H\&E staining often fails to provide sufficient diagnostic information for many diseases. Therefore, special stains are used to offer critical diagnostic insights in various dimensions ~\cite{lin2022unpaired}. For instance, in renal pathology, Masson's trichrome (MAS) stain is used to observe collagen fibers, while periodic acid-Schiff (PAS) stain is employed to examine glomerular and tubular structures. The need for multiple stains necessitates repeated tissue sampling and staining procedures, increasing labor and material costs significantly. These factors can deter patients from undergoing necessary pathological examinations, impede effective disease monitoring by clinicians, and hinder the widespread adoption of pathological diagnostics~\cite{de2021deep}.

\begin{figure}
    \centering
    \includegraphics[width=1\linewidth]{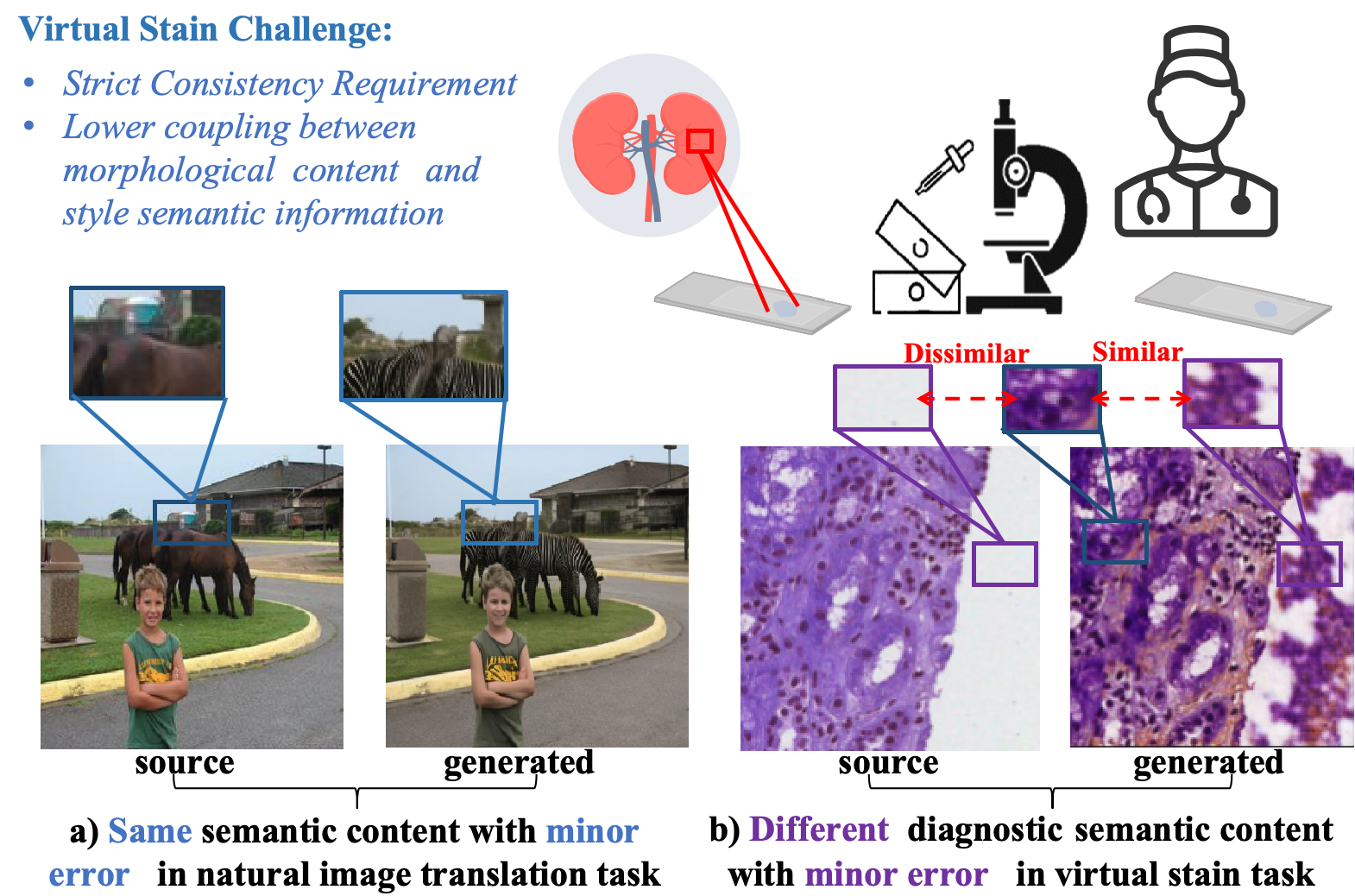}
    \caption{Structural consistency is crucial in virtual staining. a) In natural image translation, changes in non-core details do not affect the semantic content of the core. b) In pathological images, every detail holds diagnostic significance, and even minor changes can alter the diagnostic information.}
    \label{fig:overview}
\end{figure}

Recently, the rapid development of generative models has spurred interest in digital pathology, specifically in the domain of stain transformation within the same tissue sample. Unlike natural image translation where minor background changes during style transfer are often disregarded, in virtual staining, background alterations carry distinct diagnostic significance for different stains, as shown in Figure~\ref{fig:overview}. De Haan et al.  proposed a model for transferring H\&E staining to PAS, MAS, and MT stains, demonstrating the feasibility of interconversion among these non-immunochemical stains~\cite{de2021deep}. Revension et al. developed a GAN model that uses autofluorescence images to generate virtual H\&E and MT stains, affirming that tissue structure consistency is paramount in non-immunochemical staining, with different stains altering specific structure colors without affecting visible structures~\cite{rivenson2019virtual}. However, these methods rely on pixel-level paired data, which is difficult to obtain due to the physical properties of tissue samples. Chen et al. introduced an unpaired image virtual staining method, employing class one-hot labels and a learnable matrix to control the generator, enabling the virtual generation of PAS, MAS, and PASM stains~\cite{lin2022unpaired}. Zhang et al. utilized feature-content separation for mutual conversion among multiple unpaired stained pathological images~\cite{zhang2022mvfstain}. This approach, which generates various virtual stain images from any type of stained image, is known as multi-domain stain transfer.


Despite the impressive performance of GAN-based methods\cite{chen2024pathologicalsemanticspreservinglearninghetoihc,zeng2022semi,li2023adaptivesupervisedpatchnceloss,falahkheirkhah2022generative} in realistic staining style transfer, maintaining high structural consistency is challenging due to adversarial training of generators and discriminators, which is crucial for virtual staining, as the end-to-end generation nature of GANs complicates the separation of style and content~\cite{zhang2022mvfstain,li2024av,wang2024cytogan}. Overall, compared to image translation tasks in natural images, virtual staining presents unique requirements and challenges.  These include: \textbf{(1)} Stricter structural consistency requirements. In histopathological staining, staining agents color specific tissues, which means that any changes in the content can lead to deviations in pathological diagnostic semantic information. \textbf{(2)} Higher demands for the separation of content and style. Pathological images have a lower degree of coupling between morphology and semantic information, making content extraction more difficult compared to natural images. Virtual staining is a special task that assigns specific styles to specific content, and the lack of pixel-level paired data makes it challenging to completely separate style and content~\cite{zhang2022mvfstain}.

To overcome the two challenges in stain trasnfer, we first propose the pathological dual-path diffusion inversion stain transfer framework. \textit{\textbf{Firstly}}, for maintaining structual consistency, we utilize the deterministic inversion trajectory of mapping the input image to noise as one of the dual paths, called \textbf{Structual Target Path}. We optimize a visual prompt, called StainStructPrompt, around the pivtoal noise in the inversion process of mapping noise to a single image. During the process of mapping noise back to the image, we optimize null-visual prompt at each timestep to avoid the impact of cumulative inversion errors on structural consistency, rather than fine-tuning pretrained weight. \textit{\textbf{For the second challenge}}, building on the ability to recover input content, the Structural Target Path retains complete content information. We introduce a second style control path, called \textbf{Style Target Path}, to achieve fully controllable fusion of style and content. We generate images with the target staining using a feature adaptation function, which serves as the style control path in the dual-path framework. We leverage conditional sampling of the pre-trained diffusion model and null-visual prompt, called StainStylePrompt, to optimize style extraction around the style control path, thereby controlling the style. Based on the null-visual prompt optimized around pivotal noise as ``operators", our approach compensates for the inevitable minor errors in diffusion inversion. In summary,our contributions are as follows:

\begin{itemize}
    \item We are the first to handle the task of  virtual staining  using unpaired data with  a   formulation of single pretrained diffusion models, which enable us to leverage the content-style disentanglement characteristic of DDIM Inversion deterministic inversion process.
    \item We propose a dual-path prompted inversion method that employs visual prompt images to ensure precise reconstruction of both structural and style target paths. This approach maintains high structural consistency while progressively integrating target domain style.
    \item We introduce a visual prompt inversion optimization strategy that integrates target domain style information and structural information from the diffusion inversion process by optimizing null prompt images. This approach achieves disentanglement between content structural consistency and style information.
    
\end{itemize}

\section{Related Work}

\subsection{Virtual Staining in Histopathological Analysis}

In recent years, virtual staining has garnered increasing attention and rapid development in digital pathology. Researchers have applied image translation methods to virtual staining tasks, achieving remarkable results. Rivension et al. proposed a method for transferring styles between paired H\&E and PAS, MT stained images, demonstrating the learnable implicit features and associations between different staining types~\cite{rivenson2019virtual}. Zhang et al.  introduced a technique using KL loss and tissue loss to align and separate style feature spaces across different domains, achieving multi-domain unpaired data stain transfer~\cite{zhang2022mvfstain}. Lin et al.  developed a style-guided normalization algorithm for transferring unpaired H\&E to PAS, PASM, and MAS stains, exhibiting outstanding performance~\cite{lin2022unpaired}. Ma et al.  enhanced diagnostic reliability by leveraging the spatial correlation of adjacent tissue sections~\cite{ma2023agmdt}. Guan et al.  further improved the precision of style transfer between virtual stains through Renyi entropy regularization and a progressive cascade guiding block~\cite{guan2024unsupervised}. However, most of these methods focus on style transfer in virtual staining and overlook another critical principle: maintaining the structural consistency between the generated virtual stained samples and the input samples.

\subsection{Image-to-Image Translation based on Diffusion Models}

Recently, diffusion models have gained significant attention in computer vision as a novel high-fidelity generation method. However, their potential in virtual staining remains largely unexplored. Due to the likelihood evaluation characteristic of diffusion models, training with unpaired data poses substantial challenges. Some researchers, such as those behind StainDiff \cite{shen2023staindiff}, employed paired data with two parallel diffusion models to alternately fuse features from two domains. Nevertheless, this approach fails to address the unpaired data issue and requires separate training for each domain, significantly increasing computational demands. 

Another approach involves using pre-trained diffusion models with constrained guided sampling for image translation. For instance, EGSDE ~\cite{zhao2022egsde} used an energy-based guide to maintain high fidelity and consistency during sampling. Similarly, CycleNet ~\cite{xu2024cyclenet} fine-tuned large-scale pre-trained models, adding self-consistency constraints to achieve high structural consistency in unpaired image translation. However, unlike natural images, the morphological information of tissue cell structures is less coupled with semantic information. In pathological images, it is challenging to find corresponding morphological prior information. Additionally, there are no pre-trained models specifically for pathological images. Due to these reasons, diffusion models have rarely been applied to the field of virtual staining until now.

\begin{figure*}
    \centering
    \includegraphics[width=1\linewidth]{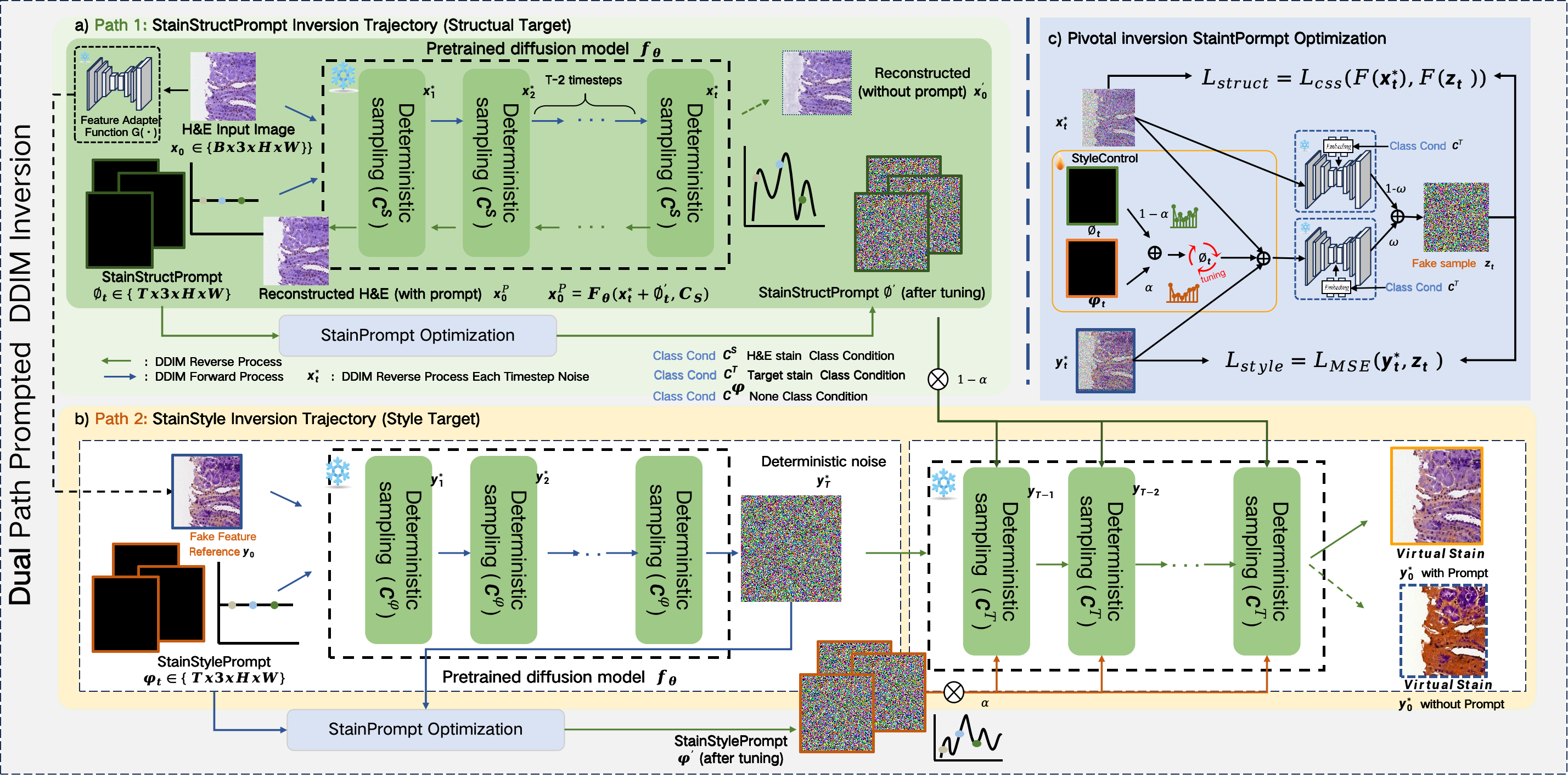}
    \caption{The overall framework of the proposed dual path inversion method. By optimize  null-visual prompt to correct deviation of deterministic sampling, Pre-trained diffusion models can maintain a high degree of structural consistency. Meanwhile, the proposed StainStyleControl leverage a constant to control the degree of influence of style trajectory.}
    \label{fig:architecture}
\end{figure*}


\section{Method}

In this section, we address two virtual staining challenges by proposing the use of Structual Target Path and Style Target Path inversion trajectories to optimize visual prompt maps. By controlling the influence between the visual prompts, we regulate the incorporation of style during the noise-to-image mapping process, as shown in Figure~\ref{fig:architecture}.

\subsection{Dual Path Prompted Strategy }

\textbf{\textit{To handle the first  challenge of strict structural consistency}}, we draw upon the concept of highly separated content and style in image editing using DDIM inversion\cite{mokady2022nulltextinversioneditingreal}. From the perspective of stain style transfer, tissue images consist of content (primary structural and morphological features) and style (stain types, including H\&E and PAS stains) ~\cite{lin2022unpaired}. We leverage the ability of DDIM Inversion to reconstruct the original image content and achieve stain transfer by incorporating the target stain style during the reverse process.

Furthermore, the method leverages a key characteristic of unconditional diffusion inversion: approximate equality~\cite{song2022denoisingdiffusionimplicitmodels}. The premise of DDIM inversion is that the forward-predicted noise and reverse-predicted noise at the same time step are approximately equal. To avoid any fine-tuning of the pre-trained model, an additional variable is introduced to act as an operational mechanism, correcting the slightly deviated feature vectors in the reverse process back to the same direction as in the forward process. Consequently, a default zero-valued feature map is used as the embedding to be optimized, termed as StainStructPrompt optimization. For each input image, only the empty visual-prompt map, referred to as \textbf{\textit{StainStructPrompt}}, is optimized. The pre-training and conditional embeddings remain unchanged. Similarly, during the process of inverting noise to image, the approach aims to gradually reduce the features of the original domain while incrementally incorporating features of the target domain~\cite{ju2023directinversionboostingdiffusionbased}. This is achieved by changing conditions during sampling to introduce target domain features. However, completely eliminating the original domain's information is challenging. \textbf{\textit{To address the challenge of semantically and controllably separating content and style}}, an additional style feature trajectory, called \textbf{\textit{StainStylePrompt}}, is introduced to optimize the visual prompt maps, functioning as a "style subtraction operator". We define input image as $x_0 \in \mathbb{R}^{N \times 3 \times H \times W}$. The two trajectories can be represented using the following equations and Algorithm~\ref{alg:algorithm}:

\begin{equation}
\begin{aligned}
     \mathbf{x^*}_{t+1} = \sqrt{\alpha_{t+1}} \left( \frac{\mathbf{x^*}_t - \sqrt{1 - \alpha_t} \boldsymbol{\epsilon}_\theta(\mathbf{x^*}_t, t,C_S)}{\sqrt{\alpha_t}} \right) \\ + \sqrt{1 - \alpha_{t+1}} \boldsymbol{\epsilon}_\theta(\mathbf{x^*}_t, t,C_S).
\end{aligned}
\end{equation}



\begin{equation}
\begin{aligned}
     \mathbf{y^*}_{t+1} = \sqrt{\alpha_{t+1}} \left( \frac{\textit{F }(\mathbf{x^*}_t) - \sqrt{1 - \alpha_t} \boldsymbol{\epsilon}_\theta(\textit{F }(\mathbf{x^*}_t), t,C_\varphi)}{\sqrt{\alpha_t}} \right) \\ + \sqrt{1 - \alpha_{t+1}} \boldsymbol{\epsilon}_\theta(\textit{F}(\mathbf{x^*}_t), t,C_\varphi).
\end{aligned}
\end{equation}
Where $F(\cdot)$ denote the feature adapter function for obtaining reference images of the target staining style. The input H\&E image is represented as $x_0$. We define the style template trajectory as $\{y^*_t\}_{t=0}^T$ and the structural target template trajectory as $\{x^*_t\}_{t=0}^T$, where $y^*_t, x^*_t \in \mathbb{R}^{N \times 3 \times H \times W}$. Here, $N$, $H$, and $W$ represent the batch size, height, and width of the image, respectively. The diffusion model's conditional variables are denoted as $C_S$ and $C_{\varphi}$, where $C_S$ represents the original domain category conditional variable, and $C_{\varphi}$ signifies the absence of additional conditional variables to mitigate errors in the style trajectory.

The feature adaptation function is employed to provide target domain style features for the StainStylePrompt. Ideally, different stains of adjacent slices would be the best samples to provide these features. However, for practical applications, we need to pre-learn this mapping relationship, as using adjacent slices would violate the original intention of virtual staining applications. In previous research, Zhang et al. proposed using a digital staining matrix to transfer staining styles to unstained autofluorescence images ~\cite{zhang2020digital}. However, the digital staining matrix requires a priori extraction from a specific dataset. For more in line with practical applications of this study, we utilize the previous state-of-art method from ~\cite{lin2022unpaired} as the feature adaptation function to generate templates with target features. The feature adaptation function can also be replaced with any method capable of generating images with target features.

In our experimental framework, we initialize the empty prompt map $\phi_T$ predicated on the inversion timesteps $T$, where $\phi_t \in \mathbb{R}^{T \times 3 \times H \times W}$.  We commence by establishing $\hat{y}_T = y_T + \phi_T$, followed by an iterative optimization procedure across the temporal domain $t = T, T-1, \ldots, 0$. This optimization process, executed for a predefined number of iterations, can be formulated as follows:


\begin{equation}
\mathop{\arg\min}\limits_{\phi} \sum^T_{t=1}||(y^*_{t-1},x^*_{t-1},y_{t-1}(\Bar{y_t}+\phi_t,C_T))||   
\end{equation}

\begin{equation}
\begin{aligned}
    \hat{y}_t=\Bar{y_t}+\phi_t
\end{aligned}
\end{equation}

\begin{equation}
\begin{aligned}
    y_{t-1}(\hat{y}_t,C_T)=\sqrt{\alpha_{t-1}} \left( \frac{\hat{y}_t - \sqrt{1 - \alpha_t} \boldsymbol{\epsilon}_\theta(\hat{y}_t, t,C_T)}{\sqrt{\alpha_t}} \right) \\
    + \sqrt{1 - \alpha_{t-1}} \boldsymbol{\epsilon}_\theta(\hat{y}_t, t,C_T).
\end{aligned}
\end{equation}
Where $\|\cdot\|$ denotes the loss function. \(y_{t-1}(\bar{y_t} + \phi_t, C_T)\) represents the use of conditional sampling with the Visual Prompt \(\phi_t\) and \(C_T\) as the category conditional encoding for the target staining domain, where \(C_T \in \{1, 2, 3\}\) corresponds to MAS, PAS, and PASM staining types. We apply the DDIM sampling step. At the end of each step, we update:

\begin{equation}
    \Bar{y}_{t-1}=y_{t-1}(\Bar{y_t}+\phi_t,C_T)
\end{equation}

This approach results in high-quality structural consistency in reconstruction, while still allowing for intuitive virtual staining by simply optimizing the prompt map in the direction of the target staining features and applying category sampling. Additionally, after a single inversion process, the same structural null-prompt embedding can be used for unpaired staining transfer by adding features to the input image. Although fine-tuning the entire model can yield more expressive results in natural image editing ~\cite{mokady2023null}, this does not apply to virtual staining in histopathology. Therefore, optimizing an ``operator" for the pivotal noise in the forward process, given the absence of large-scale pre-trained models in histopathology, is currently the most suitable solution.


\subsection{StainPrompt Optimization}


We use dual-path trajectories as templates for noise-to-image mapping. Visual prompts act as ``operators'', approximating these templates and seeking an optimal balance. We control style and content in color transfer by adjusting weights of two visual prompts with a constant $\lambda$, optimizing the Visual Prompt (StainPrompt) for both structure and staining style. When updating StainPrompt, two types of loss terms are used: structural loss $l^z_{struct}$ and stain style loss $l^y_{style}$.

\begin{equation}
    l_{StainPrompt}=\lambda* l^z_{struct}+(1-\lambda)* l^y_{style}
    \label{eq:balance}
\end{equation}

Here,we formulate $l^z_{struct}$ and $l^y_{style}$ as follow:

\begin{equation}
    l^z_{struct}=\sum_{t=0}^{IST}\frac{2(\sigma_{z_t}\sigma_{y_t}+c_1)\times(\sigma_{z_ty_t}+c_2)}{(\sigma_{z_t}^2+\sigma_{y_t}^2+c1)(\sigma_{z_t}\sigma_y+c_2)}
\end{equation}
\begin{equation}
    l^y_{style}=\sum_{t=0}^{IST}||y^*_{t}-y_{t}||_2^2
\end{equation}
Here, $\sigma_{z}$ and $\sigma_{y}$ denote the variances of the structural trajectory $z$ and the reverse sampling noise $y$, respectively, while $\sigma_{zy}$ represents the covariance between them. IST denote the number of optimization steps in StainPrompt.The value of $\alpha$ is manually configured, and its impact is discussed in the Section Experiment and Results and appendix.

We observed that errors accumulate as the reverse process propagates, and excessive iterations do not significantly improve the quality of the generated output. Therefore, to reduce time consumption, we determine the number of optimization steps for the prompt map (IST) as follows:

\begin{equation}
    IST_t=(1-\frac{t}{T})\times IST_{init}
\end{equation}
Here, $\text{IST}_t$ represents the number of iterations at the $t$-th time step, and $\text{IST}_{\text{init}}$ denotes the initially set number of iterations.

\begin{algorithm}[tb]
\caption{Dual Path Prompting Optimization Algorithm}
\label{alg:algorithm}
\textbf{Require}: Pre-trained diffusion model $\boldsymbol{f_\theta}$, Feature Adapter $\boldsymbol{F(\cdot)}$\\
\textbf{Input}: H\&E image $x_0 \in \mathbb{R}^{N \times 3 \times H \times W}$\\
\textbf{Setting}: $IST_{init}$, Timesteps $T$\\
\textbf{Output}: Generated image $y_0$
\begin{algorithmic}[1]

\STATE \textbf{Stage 1: Build Dual-Path Trajectory}
\STATE Initialize $t \gets 0$ and $y^*_0 \gets F(x_0)$
\WHILE{$t < T$}
    \STATE $\mathbf{x^*}_{t+1} \gets \sqrt{\alpha_{t+1}} \Big( 
    \frac{\mathbf{x^*}_t - \sqrt{1 - \alpha_t} \boldsymbol{f}_\theta(\mathbf{x^*}_t, t, C_s)}{\sqrt{\alpha_t}}
    \Big) + \sqrt{1 - \alpha_{t+1}} \boldsymbol{f}_\theta(\mathbf{x^*}_t, t, C_s)$
    \STATE $\mathbf{y^*}_{t+1} \gets \sqrt{\alpha_{t+1}} \Big( 
    \frac{\mathbf{y^*}_t - \sqrt{1 - \alpha_t} \boldsymbol{f}_\theta(\mathbf{y^*}_t, t, C_\varphi)}{\sqrt{\alpha_t}}
    \Big) + \sqrt{1 - \alpha_{t+1}} \boldsymbol{f}_\theta(\mathbf{y^*}_t, t, C_\varphi)$
    \STATE $t \gets t + 1$
\ENDWHILE
\STATE \textbf{return} $\{\mathbf{x^*}_t, \mathbf{y^*}_t\}_{t=0}^T$

\STATE \textbf{Stage 2: StainPrompt Optimization}
\STATE Initialize $t \gets T$, StainPrompt $\phi \in \mathbb{R}^{T \times 3 \times H \times W}$
\WHILE{$t \geq 1$}
    \FOR{$j = 0, \ldots, IST_t$}
        \STATE $\phi_t \gets \phi_t - \nabla_\phi \Big( 
        \lambda \|y_t(\bar{y}_{t-1} + \phi_t, C_T) - y^*_t\|^2 + (1 - \lambda) l^x_{struct} \Big)$
    \ENDFOR
    \STATE $t \gets t - 1$, $\bar{y}_t \gets y_t(\bar{y}_{t-1} + \phi_t, C_T)$
\ENDWHILE
\STATE \textbf{return} StainPrompt $\phi_T$

\STATE \textbf{Stage 3: Deterministic Sampling}
\STATE Initialize $t \gets T$
\WHILE{$t \geq 1$}
    \STATE $y_{t-1} \gets \sqrt{\alpha_{t-1}} \Big( 
    \frac{y_t + \phi_t - \sqrt{1 - \alpha_t} \boldsymbol{f}_\theta(y_t + \phi_t, t, C_T)}{\sqrt{\alpha_t}}
    \Big) + \sqrt{1 - \alpha_{t-1}} \boldsymbol{f}_\theta(y_t + \phi_t, t, C_T)$
    \STATE $t \gets t - 1$
\ENDWHILE
\STATE \textbf{return} $y_0$

\end{algorithmic}
\end{algorithm}

\begin{figure}
    \centering
    \includegraphics[width=0.8\linewidth]{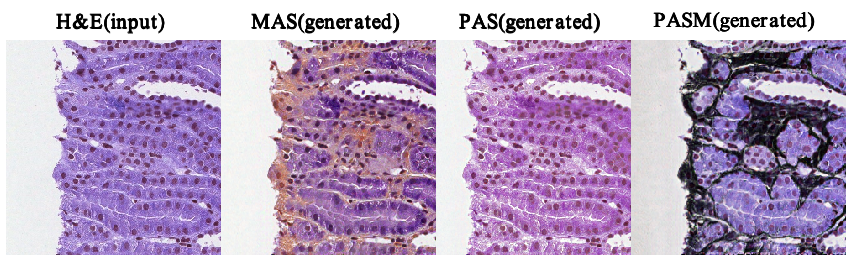}
    \caption{The virtual generation of MAS, PAS, and PASM
stained images from H\&E stained images using our network.}
    \label{fig:alldomain}
\end{figure}


\begin{figure*}
    \centering
    \includegraphics[width=0.75\linewidth]{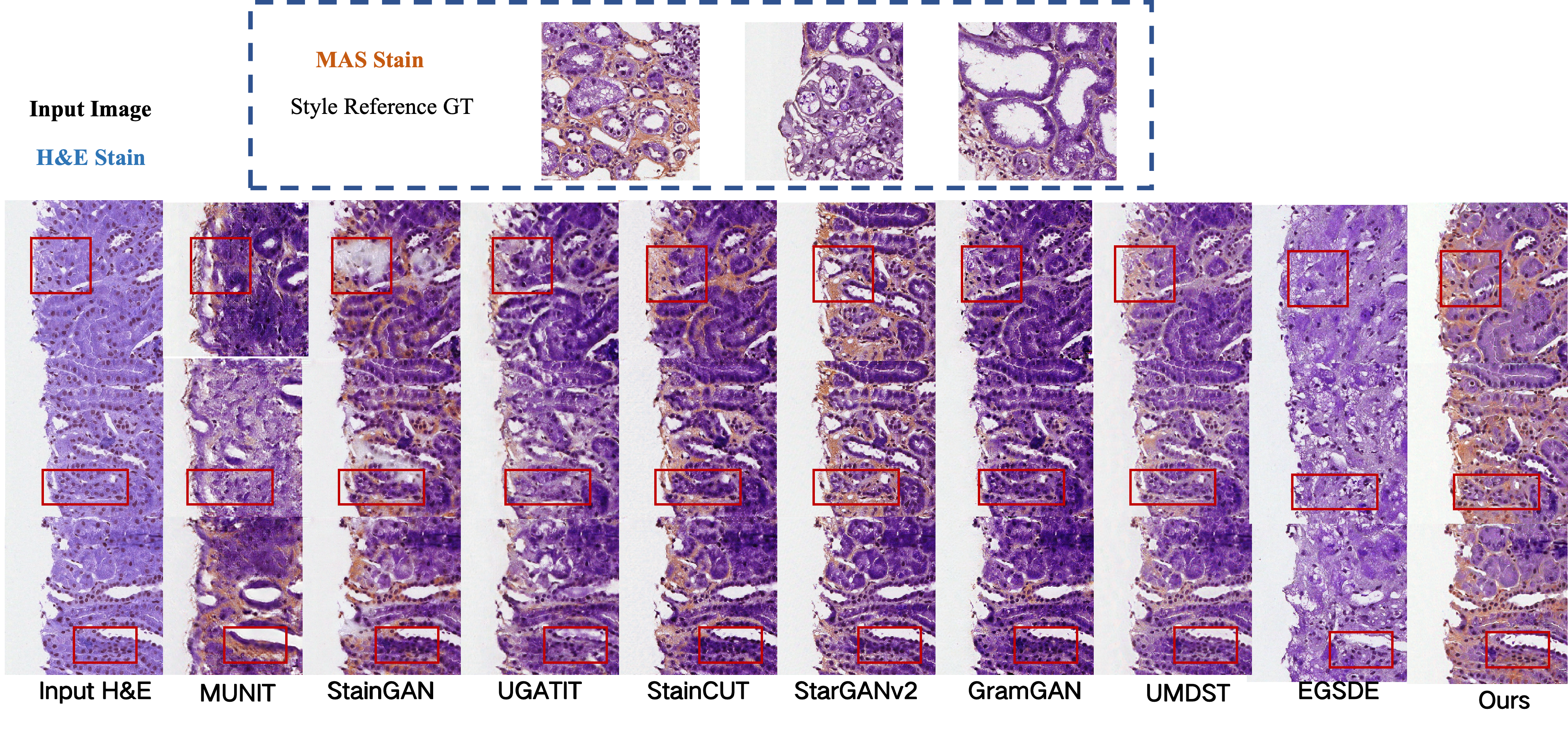}
    \caption{Comparison of different methods of staining migration from the same H\&E staining image to MAS staining. Our model performs state-of-art in various indicators and subtle structures.}
    \label{fig:mas}
\end{figure*}

\begin{table*}[ht]
\centering
\setlength{\tabcolsep}{1mm} 
\begin{tabular}{lccccccccccc}
\toprule
\textbf{Method} & \multicolumn{5}{c}{\textbf{H\&E2MAS}} & \multicolumn{5}{c}{\textbf{H\&E2PAS}} \\
\cmidrule(lr){2-6} \cmidrule(lr){7-11}
 & \textbf{SSIM$\uparrow$} & \textbf{CSS$\uparrow$} & \textbf{MS-SSIM$\uparrow$} & \textbf{PSNR$\uparrow$} & \textbf{FID} & \textbf{SSIM$\uparrow$} & \textbf{CSS$\uparrow$} & \textbf{MS-SSIM$\uparrow$} & \textbf{PSNR$\uparrow$} & \textbf{FID} \\
\midrule
MUNIT & 0.1206 & 0.1281 & 0.1705 & 10.7914 & 159.724 & 0.0758 & 0.0843 & 0.04936 & 9.8460 & 217.625 \\
StainCUT & 0.4686 & 0.5039 & 0.7603 & 14.315 & 111.751 & 0.6871 & 0.7006 & 0.8673 & 17.193 & 119.493 \\
StarGANv2 & 0.5703 & 0.6143 & 0.7448 & 14.446 & \textbf{101.679} & 0.6277 & 0.6432 & 0.7997 & 16.538 & 100.187 \\
UGATIT & 0.7110 & 0.7552 & 0.7283 & 15.563 & 127.274 & 0.7080 & 0.7260 & 0.7722 & 16.598 & \textbf{93.545} \\
StainGAN & 0.8124 & 0.8620 & 0.78139 & 16.1558 & 128.403 & 0.8455 & 0.8664 & 0.9033 & 17.5423 & 136.358 \\
GramGAN & 0.6027 & 0.6784 & 0.8001 & 13.4963 & 151.838 & 0.7167 & 0.7324 & 0.8757 & 16.989 & 147.841 \\
UMDST & 0.7329 & 0.76104 & 0.8725 & 17.559 & 179.197 & 0.7396 & 0.7616 & 0.9233 & 16.4336 & 160.589 \\
\bottomrule
EGSDE(Diffusion-based) & 0.1345 & 0.13602 & 0.3066 & 14.7350 & 197.860 & 0.1698 & 0.1712 & 0.3287 & 16.221 & 173.066 \\
\textbf{Ours} & \textbf{0.9233} & \textbf{0.9309} & \textbf{0.9458} & \textbf{20.915} & 179.289 & \textbf{0.9114} & \textbf{0.9167} & \textbf{0.9500} & \textbf{22.481} & 146.174 \\
\bottomrule
\end{tabular}
\caption{Comparison of different virtual stain methods in terms of multiple metric for both H\&E2MAS and H\&E2PAS. In the H\&E2PAS experiment, we directly selected the compromise weight($\lambda=0.55$). In the H\&E2MAS experiment, we selected struct weight($\lambda=0.05$). More experiment details and clinical disscussion are included in appendix.}.
\label{table:comparison_merged}
\end{table*}

\section{Experiment and Results}


\subsection{Dataset}

Our dataset setup follows UMDST~\cite{lin2022unpaired}, but due to the numerous patches generated from whole slide images, many lack tissue samples or contain very few, leading to different filtering criteria. In the ANHIR dataset~\cite{borovec2020anhir}, we used kidney tissue slices, each set containing four consecutive slices stained with H\&E, MAS, PAS, or PASM. Despite being pixel-level unpaired, these slices show spatial structural similarity due to their consecutive nature, though non-adjacent slices still differ significantly. We used whole slide images of kidney tissue from patients 1-4 at 40x magnification for training, and patient 5 for testing. The H\&E stained sample from patient 1 was excluded from training due to significant staining differences. Images were divided into 256x256 patches with a 192-pixel overlap. The training set includes 42,167 images (7,688 H\&E, 12,132 MAS, 11,458 PAS, and 10,889 PASM stained). The test set contains 1,989 H\&E, 2,062 MAS, 1,900 PAS, and 2,119 PASM stained images.


\subsection{Evaluation Metrics}
We evaluated our work across multiple dimensions. While extracting the morphological content of the input image is a critical step in stain transfer ~\cite{lin2022unpaired}, previous non-immunohistochemical virtual staining methods predominantly focused on measuring structural information using SSIM or assessing style using FID alone. We found that, in addition to the importance of structure and style evaluation, the peak signal-to-noise ratio (PSNR) is also crucial. We evaluate our method by SSIM, CSS, MS-SSIM, FID, PSNR.







\subsection{Experimental Details}
Our model was implemented in Python using PyTorch on an Ubuntu server with 48 GB memory and an NVIDIA A6000 GPU. During pre-training, images were randomly flipped vertically and horizontally. The diffusion model was trained for 80k iterations with a batch size of 8 and a learning rate of \(1 \times 10^{-4}\) using the ADAMW optimizer. For validation, the batch size was set to 1, and DDIM inversion timesteps were set to 100. We set the structural loss function constants \(c_1\) and \(c_2\) to \(1 \times 10^{-8}\) and \(\text{IST}_{\text{init}}\) to 50.

For feature adaptation, we followed Lin et al.~\cite{lin2022unpaired}, pre-training according to their setup. For StainPromptLoss, we determined \(\alpha\) by evaluating virtual staining on ten randomly selected validation images at intervals of 0.05 from 0 to 1. \(\alpha\) was set to 0.75 for H\&E to MAS staining and 0.55 for H\&E to PAS staining. Relevant supporting materials will be provided in the Appendix.


\subsection{Comparison Results}
Here, we compare our method with previous unpaired image translation and virtual staining methods. The baseline methods include advanced GAN-based techniques such as MUNIT\cite{huang2018munit}, UGATIT\cite{kim2020ugatitunsupervisedgenerativeattentional}, StarGAN2\cite{choi2020starganv2}, StainGAN, StainCUT~\cite{park2020cut}, GramGAN, and UMDST, and EGSDE, based on a pre-trained diffusion model. Extensive experiments demonstrate that, for the problem of multi-domain unpaired data stain transfer, our method achieves competitive results across multiple metrics, particularly in terms of structural consistency. as shown Figure~\ref{fig:mas} and Table~\ref{table:comparison_merged}.

As shown in Table~\ref{table:comparison_merged} and Figure~\ref{fig:alldomain}, our method demonstrates high structural consistency and fidelity in stain transfer results across various metrics in unpaired data multi-domain virtual stain task. Due to the choice of feature adaptation functions, our approach improves structural consistency and style transfer quality over the non-state-of-the-art baseline UMDST, achieving state-of-the-art performance across multiple metrics(SSIM,CSS,MS-SSIM,PSNR). 

As seen in Figure~\ref{fig:mas}, baseline methods like UMDST fail to color certain regions that should be stained, resulting in missing details and blurriness. Additionally, the CSS metric struggles to evaluate artifacts, as illustrated by the results of StainGAN. Through multiple metric evaluations, it is evident that our method achieves the most reliable results.




\subsection{Ablation Study}
\subsubsection{Effectiveness of StainStructPrompt}
\begin{figure}[!h]
    \centering
    \includegraphics[width=0.9\linewidth]{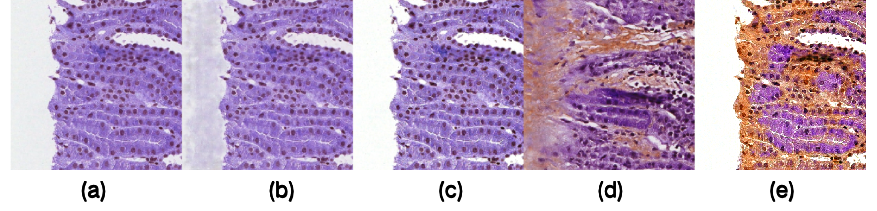}
    \caption{Quantitative comparison with StainStructPrompt: (a) Input H\&E image, (b) unconditional inversion \textbf{without} prompt, (c) unconditional inversion \textbf{with} prompt, (d) conditional inversion \textbf{without} prompt, (e) conditional inversion \textbf{with} prompt. Our structural prompt optimization tunes structural information of the deviation into the prompt.}
    \label{fig:exp_struct}
\end{figure}

As shown in Figure \ref{fig:exp_struct}, unconditional diffusion inversion can provide high structural consistency. However, the addition of conditional variables increases the content error. To validate the effectiveness of our proposed StainStructPrompt in maintaining content integrity, we examined the role of using only the structural trajectory branch on content preservation. This analysis demonstrates that the optimization of prompt structure allows for error-free inversion of the reverse trajectory \(z^*_T, \ldots, z^*_0\) of the input image in DDIM Inversion.

\begin{table}[h!]
  \centering
  \begin{tabular}{|c|c|c|c|}
    \hline
    \textbf{UDM} & \textbf{UDM (w prompt)} & \textbf{CDM} & \textbf{CDM (w prompt)} \\
    \hline
    95.2\% & \textbf{96.58\%} & 65.23\% & \textbf{81.84\%} \\
    \hline
  \end{tabular}
  \caption{The evaluation SSIM(Structual Metric,higher is better) results of Inversion with StainStructPrompt.}
  \label{tab:comparison}
\end{table}

However, as observed in Figure \ref{fig:exp_struct}, the true target domain's feature characteristics are compromised when only structural trajectory are used in conditional inversion. This issue arises because, during the optimization of pivotal noise inversion, the structural information from viusal prompt influences the overall style feature distribution through class-conditional sampling, causing a deviation in the direction of the style features. To correct the direction of these style features, we introduce a style trajectory \(y^*_T, \ldots, y^*_0\) for style feature correction while preserverving structural prompt. we balance the weight of structure and style as description equation \eqref{eq:balance}.

\begin{figure}[!h]
    \centering
    \includegraphics[width=1\linewidth]{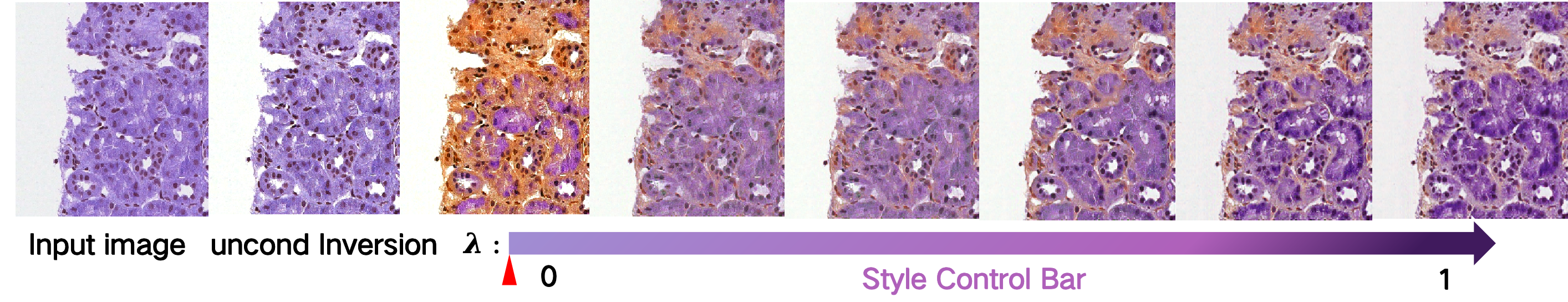}
    \caption{StainStyleControl achieves style manipulation by adjusting the weights of two prompts.}
    \label{fig:style_control}
\end{figure}

\subsubsection{Effectiveness of StainStylePrompt}

We have validated that, within our framework, prioritizing the StainStylePrompt results in a significant reduction in the Fréchet Inception Distance (FID) and an improvement in structural consistency compared to the baseline, as shown in Table~\ref{table:StylePrompt}. Besides, as shown in Figure~\ref{fig:style_control}, we can control the style by a constant $\lambda$, which means our method can separate content and style completely.

\begin{table}[ht]
\centering
\begin{tabular}{lccc}
\toprule
\textbf{Experiment} & \textbf{SSIM$\uparrow$} & \textbf{FID$\downarrow$} \\
\midrule
baseline & 0.8015 & 268.491 \\
baseline+\textbf{StainStylePrompt} & \textbf{0.8121} & \textbf{255.261}  \\
\bottomrule
\end{tabular}
\caption{Evaluation of effectiveness of StainStylePrompt using FID, SSIM.}
\label{table:StylePrompt}
\end{table}

\subsubsection{Effectiveness of Dual-Path Prompting Optimization Framework}
We demonstrate that as the StainStylePrompt becomes dominant, the style quality metric FID improves. Similarly, when the StainStructPrompt is dominant, the structural metric, Structural Similarity Index Measure (SSIM), shows better results. We have selected a balanced parameter for the stain transfer outcome, which surpasses the baseline across all evaluated metrics, as shown in Table~\ref{table:comparisonStyle}. More details regarding parameter settings and variations are discussed in the appendix.

\begin{table}[ht]
\centering
\begin{tabular}{lccc}
\toprule
\textbf{Experiment} & \textbf{SSIM} & \textbf{FID} & \textbf{PSNR} \\
\midrule
baseline & 0.8015 & 268.491 & 19.746 \\
StructPrompt domainated & \textbf{0.909}  & 278.253 & \textbf{20.867} \\
Dual-path Prompt(\textbf{slected}) & 0.8651 & 263.039 & 19.862 \\
StylePrompt domainated & 0.8474 & \textbf{246.999} & 19.833 \\
\bottomrule
\end{tabular}
\caption{Ablation study of our method. Utilizing both trajectories improves all metrics over the baseline. }
\label{table:comparisonStyle}

\end{table}

\subsubsection{Different Feature Adapter Function}

\begin{table}[ht]
\centering
\begin{tabular}{lccc}
\toprule
\textbf{Experiment} & \textbf{SSIM$\uparrow$} & \textbf{FID$\downarrow$} & \textbf{PSNR$\uparrow$} \\
\midrule
Function1 & 0.7723 & 264.375 & 17.633 \\
\textbf{Ours(G($\cdot$)=Function1)} & \textbf{0.9261} & \textbf{234.074} & \textbf{23.681} \\
\bottomrule
Function2 & 0.8789 & 238.736 & 19.032 \\
\textbf{Ours(G($\cdot$)=Function2)} & \textbf{0.9661} & \textbf{238.727} & \textbf{26.305} \\
\bottomrule
\end{tabular}
\caption{Comparison of different feature adapter functions within our framework for the H\&E to MAS stain transfer task. Our method consistently outperforms the baseline across various feature adaptation functions.
} 
\label{table:ablation}
\end{table}


As shown in the  Table \ref{table:ablation}, our method consistently produces reliable results that surpass the baseline across multiple metrics with various feature adaptation functions. We denote UMDST as Feature Adapter Function 1 (noted as Function1) and StainGAN as Feature Adapter Function 2 (noted as Function2). Experimental results demonstrate that our method exhibits strong scalability. Further, we also discuss the impact of the setting of \textit{$IST_{init}$} on the results, as shown in Appendix. For all experiments, we chose 50 as the parameter value.





\section{Conclusion}
In this paper, we propose a novel and training-free unpaired multi-domain stain transfer method based on single pretrained diffusion models.  Our method offers a solution to the structural consistency issue in virtual staining. 

\section{Acknowledgement}
This work was supported in part by the Ministry of Science and Technology’s key research and development program (2023YFF0723400), Shenzhen-Hong Kong Joint Lab on Intelligence Computational Analysis for Tumor lmaging (E3G111), and the Youth Innovation Promotion Association CAS (2022365). 

\bibliography{aaai25}

\section{Background and Preliminaries}
Denoising Diffusion Implicit Models (DDIM) are a class of generative models that can generate high-quality samples more efficiently. The key idea behind DDIM is to leverage an implicit sampling technique to reduce the number of required denoising steps while maintaining sample quality.





\subsection{Implicit Sampling in DDIM}

Unlike DDPM, which relies on a Markovian process, DDIM employs an implicit non-Markovian process for generating samples. This allows for faster sampling by skipping certain steps in the reverse process. The key formula for DDIM's reverse process is given by:

\begin{equation}
\begin{aligned}
\mathbf{x}_{t-1} = \sqrt{\alpha_{t-1}} \left( \frac{\mathbf{x}_t - \sqrt{1 - \alpha_t} \boldsymbol{\epsilon}_\theta(\mathbf{x}_t, t)}{\sqrt{\alpha_t}} \right) \\ + \sqrt{1 - \alpha_{t-1}} \boldsymbol{\epsilon}_\theta(\mathbf{x}_t, t),
\end{aligned}
\end{equation}

where $\boldsymbol{\epsilon}_\theta$ is the neural network trained to predict the noise added at each step.

\subsection{DDIM Inversion}

 The inversion process involves running the DDIM sampling process in reverse, which can be formulated as:

\begin{equation}
\begin{aligned}
\mathbf{x}_{t+1} = \sqrt{\alpha_{t+1}} \left( \frac{\mathbf{x}_t - \sqrt{1 - \alpha_t} \boldsymbol{\epsilon}_\theta(\mathbf{x}_t, t)}{\sqrt{\alpha_t}} \right) \\ + \sqrt{1 - \alpha_{t+1}} \boldsymbol{\epsilon}_\theta(\mathbf{x}_t, t).
\end{aligned}
\end{equation}

We denote $\epsilon^*_t$ as the groundtruth of prediction.To enhance the inversion process, consider the following modifications:

\begin{equation}
\mathbf{x}_0 = \mathbf{x}^*_t + \left( \frac{1}{\alpha_t} - 1 \right) \sigma, \quad \sigma > 0,
\end{equation}

\begin{equation}
\boldsymbol{\epsilon}_\theta(\mathbf{x}_t, t) = \epsilon^*_\theta(\mathbf{x}_t, t) + \sigma, \quad \sigma > 0.
\end{equation}

As $t$ increases, the error decreases. Therefore, in the non-conditional DDIM inversion process, larger time steps are typically required to reduce the error.By iteratively applying these equations, one can trace back the sample to its original noise vector, effectively performing the inversion.

\begin{figure}
    \centering
    \includegraphics[width=1\linewidth]{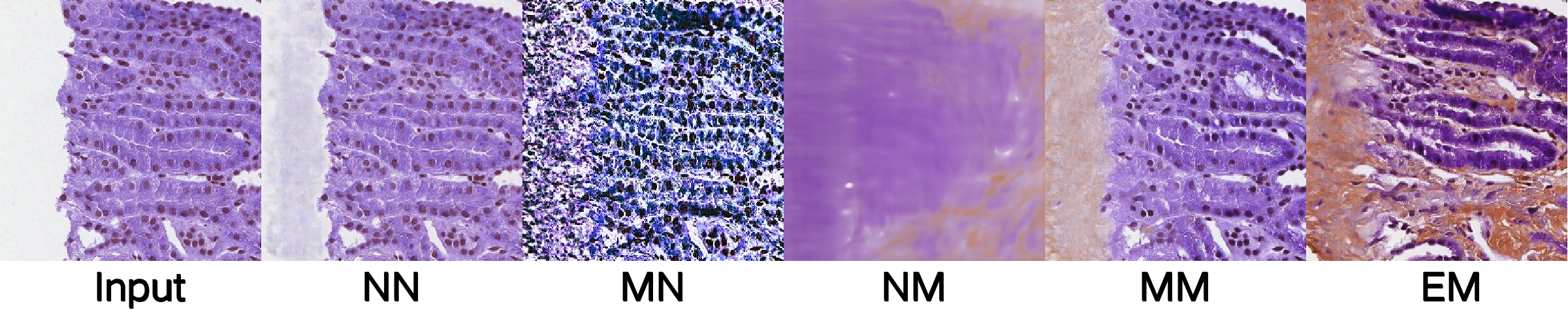}
    \caption{In the same pre-trained diffusion model, different conditional samplings affect the same input differently. The first letter represents the condition for the reverse process, and the second letter represents the condition for the forward process. N denotes no condition, M denotes the target stain domain condition, and E denotes the original domain condition.}
    \label{fig:cond}
\end{figure}

\section{More Experiment Details}
\subsection{Difference of Dual Path  Condition}
Previous work has indicated that embedding class information in the training of diffusion models effectively fits the class-conditional distribution. However, this assumes that the input images fully conform to the true data distribution. Since our dual-pathway approach leverages style reference images obtained through feature adaptation functions, the distribution of these reference images may not align with the true style distribution. Therefore, the impact of style trajectories obtained with different class conditions varies. 

As shown in Figure~\ref{fig:cond}, unconditional diffusion inversion can reconstruct complete, realistic content but introduces errors that increase with the addition of conditional variables. For a single pre-trained diffusion model, using only class condition variables can achieve target stain styles matching the true distribution, as EM in Figure~\ref{fig:cond}, but structural consistency is significantly lost.

\subsection{Parament Setting}
In our method, key components such as StainPrompt Optimization include manually set parameters, which have a significant impact on the final virtual staining results. Therefore, we studied the effects of different parameters. As shown in Table~\ref{table:istablation}, optimizing the null prompt map at each time step increases runtime. To balance runtime and performance, we chose 50 as the IST\_init parameter for this experiment.

\begin{figure}
    \centering
    \includegraphics[width=1\linewidth]{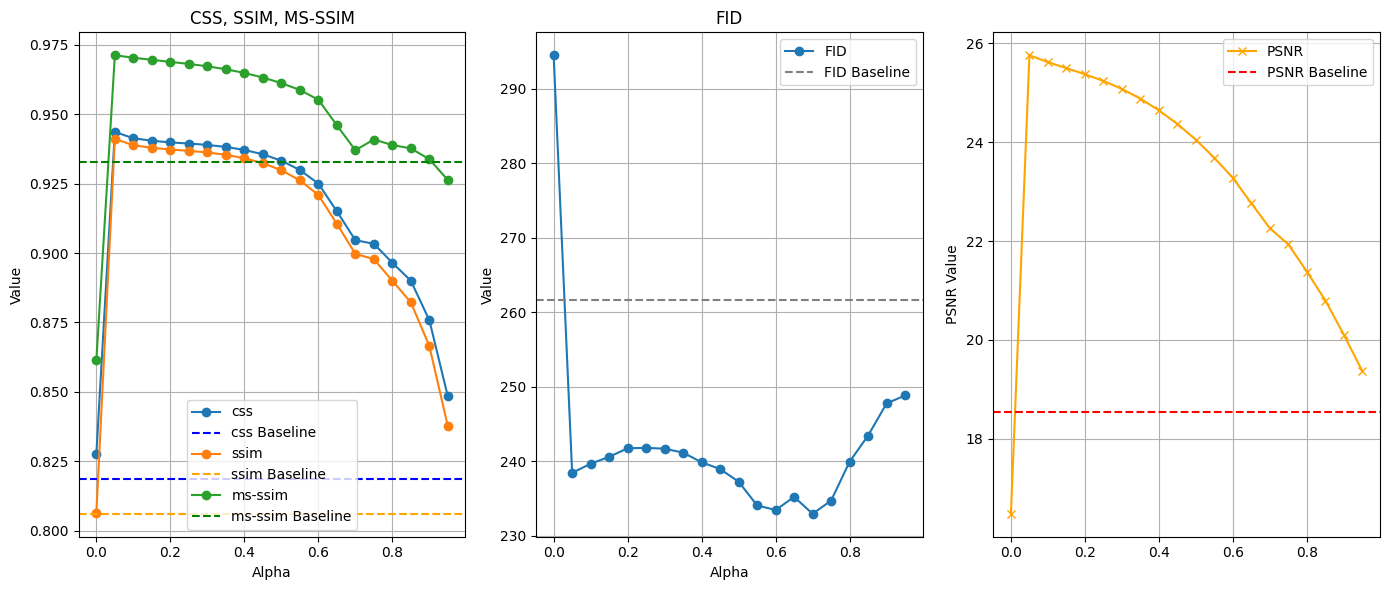}
    \caption{In the H\&E2PAS experiment, different $\lambda$ values show varying performance on different metrics. In this study, we selected $\lambda=0.55$ }
    \label{fig:pas}
\end{figure}

To avoid the influence of conditional variables from the pre-trained model on the style trajectory, we do not use conditional variables in the style trajectory. For the structure trajectory, we use conditional variables from the source domain, and for the sampling process, we use conditional variables from the target domain. As shown in Figure~\ref{fig:cond}, the correct embedding of conditional variables can achieve good style results. Based on this conditional variable embedding, we tested multiple lambda values to evaluate their impact on the results of virtual staining, as shown in the Figure~\ref{fig:mas} and Figure~\ref{fig:pas}.

\begin{table}[ht]
\centering
\begin{tabular}{lccc}
\toprule
\textbf{$IST_{init}$ setting} & \textbf{SSIM} & \textbf{FID} & \textbf{PSNR} \\
\midrule
50 & 0.9285 & 251.487 & 22.162 \\
100 & 0.9256 & 250.591 & 22.059 \\
150 & 0.9247  & 249.776 & 22.038 \\
200 & 0.9243 & 249.010 & 22.025 \\
250 & 0.9241 & 248.789 & 22.018 \\
\bottomrule
\end{tabular}
\caption{\footnotesize Different \(\text{IST}_{\text{init}}\) settings have minimal impact on the results. Therefore, we chose 50 as the experimental setting to measure time.} 
\label{table:istablation}
\end{table}

\begin{figure}
    \centering
    \includegraphics[width=1\linewidth]{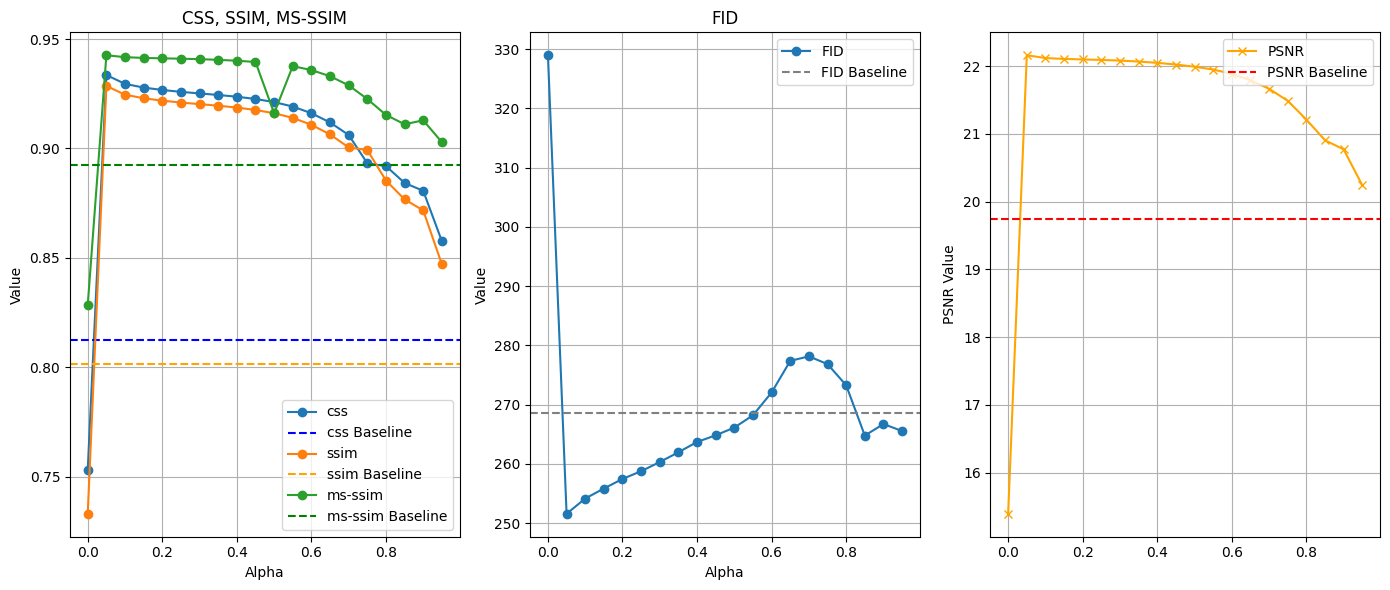}
    \caption{In the H\&E2MAS experiment, different $\lambda$ values show varying performance on different metrics. In this study, we selected $\lambda=0.05$.}
    \label{fig:mas}
\end{figure}

\begin{table}[ht]
\centering
\begin{tabular}{lccc}
\toprule
\textbf{Experiment} & \textbf{SSIM} & \textbf{FID} & \textbf{PSNR} \\
\midrule
baseline & 0.8015 & 268.491 & 19.746 \\
$\lambda=0$ & 0.699 & 302.929 & 14.086 \\
$\lambda=0.05$ & \textbf{0.909}  & 278.253 & \textbf{20.867} \\
$\lambda=0.25$ & 0.9002 & 280.730 & 20.8184 \\
$\lambda=0.45$ & 0.8863 & 278.305 & 20.2810 \\
$\lambda=0.65$  & 0.8651 & 263.039 & 19.862 \\
$\lambda=0.85$  & 0.8474 & \textbf{246.999} & 19.833 \\
$\lambda=0.95$ & 0.8121 & 255.261 & 19.220 \\
\bottomrule
\end{tabular}
\caption{\footnotesize  Comparison of methods. Utilizing both trajectories improves all metrics over
the baseline. As the weight of one trajectory increases, its
characteristic becomes more pronounced. A larger $\lambda$ value
improves style, while a smaller $\lambda$ enhances
structural integrity and overall image quality }
\label{table:comparisonStyle}
\end{table}

\section{Limitation}

\begin{figure}[h]
    \centering
    \includegraphics[width=\linewidth]{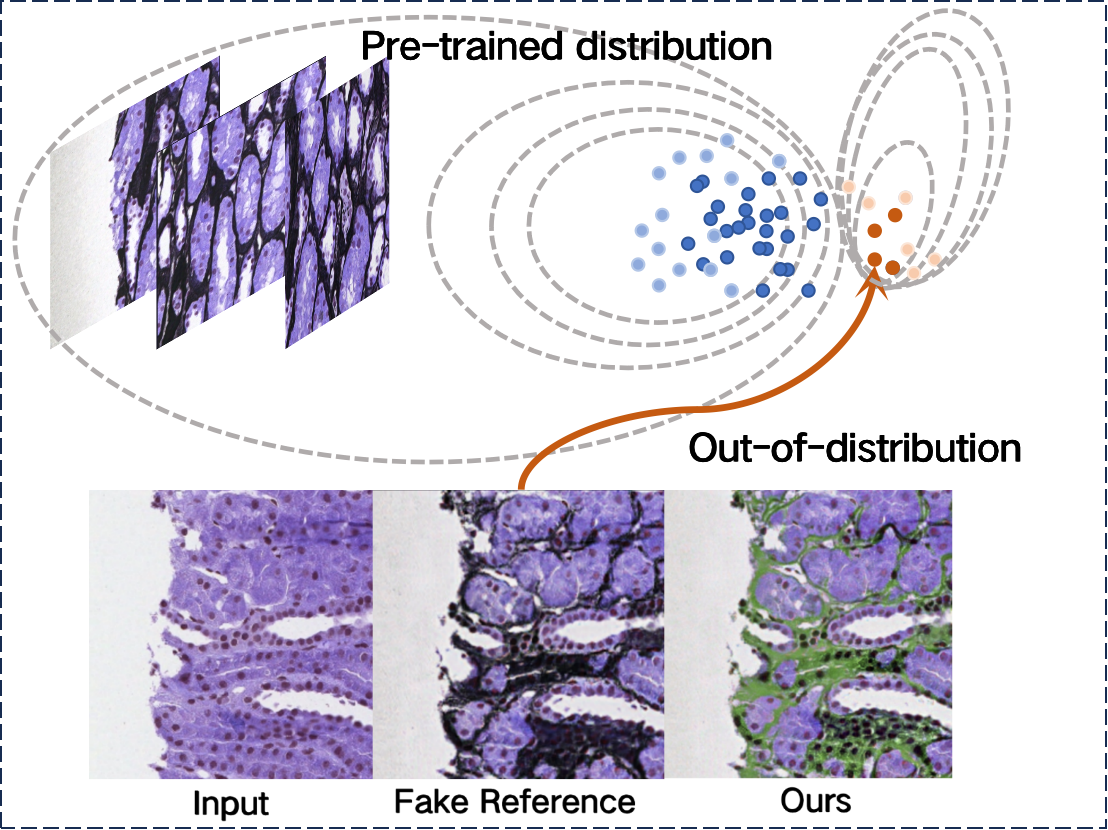}
    \caption{\footnotesize Our method exhibits failure cases in style correction under certain parameters. }
    \label{fig:limitation}
\end{figure}

When there is a substantial gap between the stains generated by the Feature Adaptation Function and the prior knowledge of true pre-trained stains, conditional encoded sampling can cause a gradual deviation in style features. As shown in the Figure~\ref{fig:limitation}, when the reference stained image incorrectly colors regions that would not be stained in reality, the StainPrompt optimization introduces style artifacts. This also demonstrates that our method maintains a high degree of structural consistency even when the style is incorrect, highlighting the strong separation between content and style in our approach.

\begin{table}[h!]
  \centering
  \begin{tabular}{|c|c|c|c|c|c|}
    \hline
    \textbf{ } & \textbf{MU} & \textbf{Cyc} & \textbf{UMD}  & \textbf{Gram}  & \textbf{ours} \\
    \hline
    ER\% & 2.76 & 4.64 & 7.96 & 6.6 & \textbf{8.36} \\
    \hline
  \end{tabular}
  \caption{The evaluation results (ER) of different models (The mean results of the 5 pathologists, 1-lowest, 10-highest).MU represent MUNIT, Cyc represent CycleGAN, UMD represent UMDST(AAAI'22), Gram represent GramGAN(TMI'24)}
  \label{tab:comparison}
\end{table}

\end{document}